# Stable two-dimensional solitons in nonlinear lattices

Yaroslav V. Kartashov,[1] Boris A. Malomed,[2] Victor A. Vysloukh,[3] and Lluis Torner[1]

[1]ICFO-Institut de Ciencies Fotoniques, and Universitat Politecnica de Catalunya, Mediterranean Technology Park, 08860 Castelldefels (Barcelona), Spain

[2]Department of Physical Electronics, School of Electrical Engineering, Faculty of Engineering, Tel Aviv University, Tel Aviv, 69978, Israel

[3]Departamento de Fisica y Matematicas, Universidad de las Americas – Puebla, Santa Catarina Martir, 72820, Puebla, Mexico

We address the existence and stability of two-dimensional solitons in optical or matter-wave media, which are supported by purely nonlinear lattices in the form of a periodic array of cylinders with self-focusing nonlinearity, embedded into a linear material. We show that such lattices can stabilize two-dimensional solitons against collapse. We also found that stable multipoles and vortex solitons are also supported by the nonlinear lattices, provided that the nonlinearity exhibits saturation.

Creation of stable two-dimensional (2D) solitons and vortex solitons in optical media and Bose-Einstein condensates (BECs) with cubic nonlinearity is a well-known challenge, because collapse as well as azimuthal instability of vortices are inherent features of 2D settings [1]. It was shown experimentally in photorefractive crystals [2-4] and predicted for cubic (Kerr) media [5-8] that the stabilization of 2D solitons and vortices may be provided by periodic potentials (*optical lattices*). A different stabilization mechanism makes use of *nonlinear lattices* [9], i.e., a spatially transverse periodic modulation of the nonlinearity. In BECs such a setting is achieved by means of Feshbach resonances induced by nonuniform magnetic fields [10]. In Optics the concept may be implemented with suitable combinations of different materials. Solitons supported by one-dimensional nonlinear lattices [11-13], as well as by combinations of linear and nonlinear ones [14-16], have been studied in detail. However, stabilization of 2D solitons and vortices in nonlinear lattices remains a challenge. Attempts to find stable solitons in nonlinear radial lattices defined similar to their linear counterparts [17,18], have failed, the only positive result reported thus far being stable axisymmetric solitons supported by a cylinder with focusing nonlinearity, embedded in a linear or defocusing nonlinear medium [19].

In this Letter we report the existence of stable 2D solitons in a lattice composed of focusing cylinders surrounded by a linear medium with the same refractive index, i.e., a purely nonlinear periodic lattice. We show that Kerr lattices of this type support stable fundamental solitons, while stability of multipoles and vortices requires saturation of the nonlinearity. Both the mechanism and the results are different from those encountered in models with linear lattices. Note that a crucial difference between linear and nonlinear lattices is their nature at low powers, where the former behave as a medium with a periodic refractive index while the latter remains a homogeneous medium. This has both practical and fundamental implications, in particular solitons in nonlinear lattices do not bifurcate from linear Bloch waves and thus exhibit different cutoff properties, as described below.

The evolution of a light beam or matter-wave packet in a medium with a nonlinear lattice is governed by the nonlinear Schrödinger equation for field amplitude $q$,

$$i\frac{\partial q}{\partial \xi} = -\frac{1}{2}\left(\frac{\partial^2 q}{\partial \eta^2} + \frac{\partial^2 q}{\partial \zeta^2}\right) + \sigma(\eta,\zeta)\frac{q|q|^2}{1+S|q|^2},\tag{1}$$

where $\xi$ is the propagation distance in optics or time in the case of BEC and $\eta,\zeta$ the transverse coordinates. The local nonlinearity coefficient is $\sigma = -1$ inside each cylinder of radius $w_r$, and $\sigma = 0$ between them, while $S \geq 0$ accounts for the saturation of the focusing nonlinearity. The cylinders form a square array with spacing $w_s$ along axes $\eta$ and $\zeta$. We fix the transverse scales by setting $w_r = 1$, and vary spacing $w_s$. Obviously, $w_s > 2$ and $\sqrt{2} \leq w_s \leq 2$ pertain, respectively, to separated cylinders and partly overlapping ones, the limiting case $w_s = \sqrt{2}$ corresponds to the uniform focusing medium.

We search for soliton solutions to Eq. (1) with propagation constant/chemical potential $b$ as $q = w(\eta,\zeta)\exp(ib\xi)$, where $w$ is real for fundamental solitons or multipoles, and complex for vortices. For the linear stability analysis, perturbed solutions were taken as $q = [w + u\exp(\delta\xi) + iv\exp(\delta\xi)]\exp(ib\xi)$. In the case of real $w$, this substitution leads to a linear problem for perturbation modes $(u,v)$ and eigenvalues $\delta = \delta_r + i\delta_i$.

$$\delta u = -\frac{1}{2}\left(\frac{\partial^2 v}{\partial \eta^2} + \frac{\partial^2 v}{\partial \zeta^2}\right) + bv + \sigma(\eta,\zeta)v\frac{w^2 + Sw^4}{(1 + Sw^2)^2},$$

$$\delta v = +\frac{1}{2}\left(\frac{\partial^2 u}{\partial \eta^2} + \frac{\partial^2 u}{\partial \zeta^2}\right) - bu - \sigma(\eta,\zeta)u\frac{3w^2 + Sw^4}{(1 + Sw^2)^2},$$

(2)

Generic examples of fundamental solitons supported by the nonlinear lattice are shown in Fig. 1, and properties of their families are summarized in Fig. 2. In contrast to uniform cubic media $(\sigma \equiv -1, S = 0)$, where for the unstable Townes soliton one has $U_T \approx 5.85$ [the dashed line in Figs. 2(a,b)], the power of solitons supported by the nonlinear lattice is a non-monotonous function of $b$ [Fig. 2(a)]. The power rapidly grows for $b \to 0$, as the soliton expands across the lattice [Fig. 1(a)]. Note, however, that, in contrast to solitons in linear lattices bifurcating from amplitude-modulated Bloch waves, low-power solitons in nonlinear lattices remain almost unmodulated. Increasing $b$ results in confinement of the soliton to a single site of the nonlinear lattice [Fig. 1(b)], a localization that is accompanied by a change of the sign of slope $dU/db$ of the $U(b)$ curve. Thus, solitons in the nonlinear lattice exist above a minimum power, $U_m$. This value decreases with increasing spacing $w_s$ between the cylinders, approaching its minimum at $w_s = \infty$, which corresponds to the soliton supported by a single cylinder [Fig. 2(b)].

The non-monotonous dependence $U(b)$ suggests that the nonlinear lattice may stabilize the fundamental solitons against collapse, in accordance with the Vakhitov-Kolokolov (VK) criterion. Indeed, a linear-stability analysis based on the numerical solution of Eqs. (2) shows that the part of the fundamental-soliton family with $dU/db > 0$ is linearly stable [Fig. 2(c) shows that $\delta_r$ vanishes exactly at the point where $dU/db$ changes its sign]. In terms of $b$, the instability domain located near the low-amplitude cutoff is narrowest when $w_s \to \infty$, and, as expected on intuitive grounds it expands as $w_s \to \sqrt{2}$, as solitons are unstable in 2D uniform cubic media. These findings represent the first example of stable 2D solitons in a purely nonlinear lattice. Inclusion of nonlinearity saturation $(S > 0)$ results in a substantial expansion of the stability domain for the fundamental solitons even at $w_s \to \sqrt{2}$, as collapse is absent even in uniform media with $S > 0$. At $S > 0$ there also exists an upper cutoff, where the soliton power diverges [Fig. 2(d)]. This cutoff is chiefly determined by the radius of the cylinders forming the nonlinear lattice.

Besides fundamental solitons, the nonlinear lattice also supports multipole and vortex states. Generic examples of such solitons are shown in Figs. 1(c)-1(f). In particular, for dipoles the mutual repulsion between out-of-phase constituents is compensated by their trapping in the self-induced nonlinear lattice. One can see [Fig. 1(c)] that the maxima of the constituents forming the dipole are slightly shifted from centers of the cylinders where they are trapped. This effect is almost negligible at high powers [Fig. 1(d)].

The simplest vortex solitons in nonlinear lattices, that we call *off-site* states by analogy with linear lattices, as they place the center between nonlinear cylinders, feature four intensity peaks [Figs. 1(e,f)], and expand across the lattice in the small-amplitude limit. We have found that, with the cubic nonlinearity, all dipoles, multipoles, and vortices are unstable, but they are readily stabilized by the saturation of the nonlinearity. Typical $U(b)$ dependencies for dipole and vortex solitons are shown in Fig. 3(a). Both types feature the threshold (minimum) power necessary for their existence (naturally, the minimum power for vortices exceeds that for dipoles, which, in turn, is higher than for the fundamental solitons). The upper cutoff of the existence domain, where the power diverges, is identical for all types of the solitons, as it is determined by the divergence of the power of the fundamental solitons, including those which form the complexes. Domains of oscillatory instabilities, corresponding to complex $\delta$, were found for both dipoles and vortices in the low-amplitude limit [see Figs. 3(b,c) for the corresponding $\delta_r(b)$ dependencies], when, as said above, they expand across the lattice [Figs. 1(c,e)]. As the localization improves, the dipoles and vortices get stable above a critical value of $b$. At equal values of $w_s$, the width of the instability domain for the vortices is substantially larger than for the dipoles. For both types, the instability domain rapidly shrinks with the increase of $w_s$, as a larger separation between cylinders results in weaker interactions between constituents forming the complexes. Other types of higher-order soliton states, such as on-site vortices and quadrupoles, may also be stable in the saturable nonlinear lattice.

We conclude by stressing the important differences between solitons supported by purely nonlinear and linear (or combined) lattices. Since nonlinear lattices do not feature a band-gap spectrum low-power solitons do not transform into Bloch waves, thus their cutoff is always $b = 0$. Such cutoff is identical for both fundamental and higher-order solitons in contrast to linear lattices. Also, while in linear lattices made in media with saturable nonlinearity, high-power solitons expand across the lattice exhibiting multiple oscillations, in the nonlinear lattice addressed here high-power solitons remain strongly confined inside the nonlinear domains. Finally, in the nonlinear lattices at high powers solitons do not undergo splitting or any other power-dependent shape transformations that are typical of combined lattices [15,16].

# References with titles

# References without titles

**Figure captions**

Figure 1.    Profiles of the absolute value of the field for fundamental solitons [(a) and (b)], dipoles [(c) and (d)], and off-site vortices [(e) and (f)]. Panels (a), (c), and (e) correspond to $b = 0.05$, while (b), (d), and (f) - to $b = 7$. Here and in Figs. 2(d) and 3, $w_\mathrm{s} = 3$ and $S = 0.1$.

Figure 2.    (a) The soliton's power versus $b$ for several values of $w_\mathrm{s}$ in the cubic medium. (b) The minimum power versus lattice spacing $w_\mathrm{s}$. The horizontal dashed line in (a) and (b) corresponds to the Townes-soliton power, $U_\mathrm{T} \approx 5.85$. (c) The real part of the perturbation growth rate versus $b$ at $w_\mathrm{s} = 3$. (d) The power versus $b$ in the saturable medium. Circles in (d) correspond to solitons from Figs. 1(a) and 1(b).

Figure 3.    (a) The power versus $b$ for dipoles and vortices. The real part of the perturbation growth rate versus $b$ for dipoles (b) and vortices (c). Circles in (a) corresponds to solitons from Figs. 1(c)-1(f).

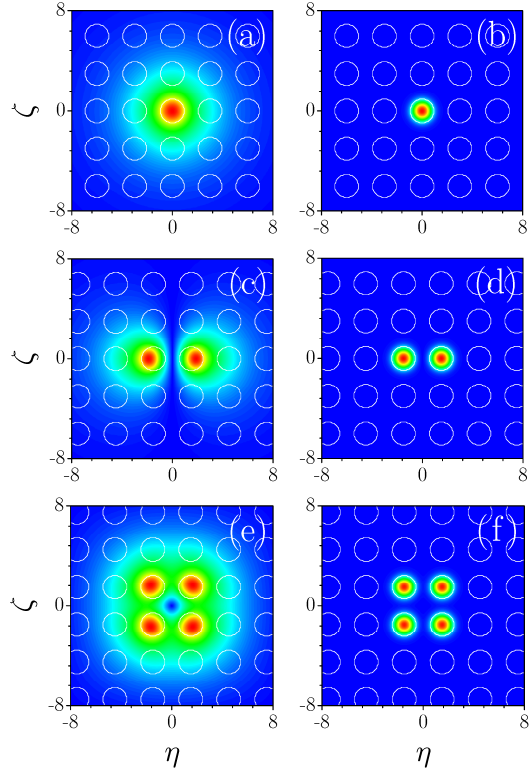

Figure 1.    Profiles of the absolute value of the field for fundamental solitons [(a) and (b)], dipoles [(c) and (d)], and off-site vortices [(e) and (f)]. Panels (a), (c), and (e) correspond to $b = 0.05$, while (b), (d), and (f) - to $b = 7$. Here and in Figs. 2(d) and 3, $w_s = 3$ and $S = 0.1$.

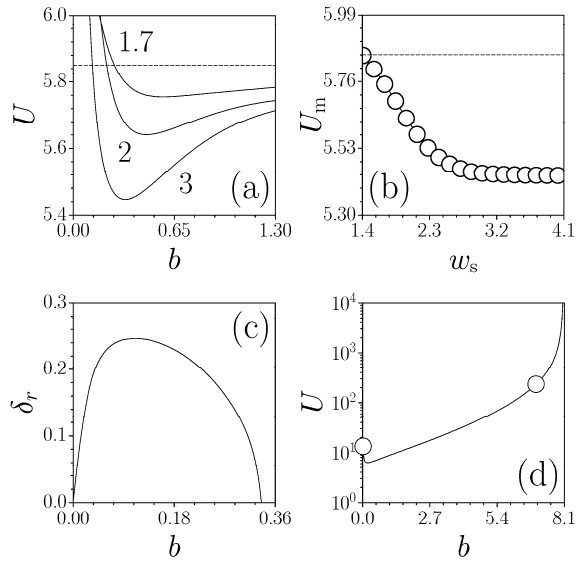

Figure 2. (a) The soliton's power versus $b$ for several values of $w_s$ in the cubic medium. (b) The minimum power versus lattice spacing $w_s$. The horizontal dashed line in (a) and (b) corresponds to the Townes-soliton power, $U_T \approx 5.85$. (c) The real part of the perturbation growth rate versus $b$ at $w_s = 3$. (d) The power versus $b$ in the saturable medium. Circles in (d) correspond to solitons from Figs. 1(a) and 1(b).

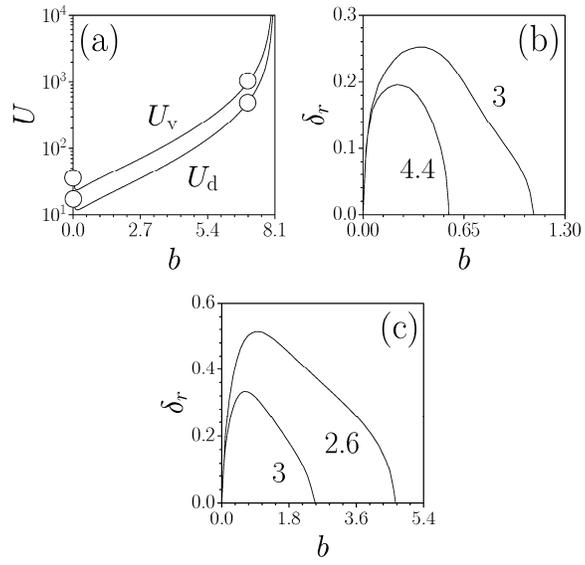

Figure 3.    (a) The power versus $b$ for dipoles and vortices. The real part of the pertur-
bation growth rate versus $b$ for dipoles (b) and vortices (c). Circles in (a) cor-
responds to solitons from Figs. 1(c)-1(f).